# Existence of Multiagent Equilibria with Limited Agents

**Michael Bowling**                                    BOWLING@CS.UALBERTA.CA
*Department of Computing Science, University of Alberta*
*Edmonton, Alberta, Canada, T6G 2E8*

**Manuela Veloso**                                     MMV@CS.CMU.EDU
*Computer Science Department, Carnegie Mellon University*
*Pittsburgh PA, 15213-3891*

## Abstract

Multiagent learning is a necessary yet challenging problem as multiagent systems become more prevalent and environments become more dynamic. Much of the groundbreaking work in this area draws on notable results from game theory, in particular, the concept of Nash equilibria. Learners that directly learn an equilibrium obviously rely on their existence. Learners that instead seek to play optimally with respect to the other players also depend upon equilibria since equilibria are fixed points for learning. From another perspective, agents with limitations are real and common. These may be undesired physical limitations as well as self-imposed rational limitations, such as abstraction and approximation techniques, used to make learning tractable. This article explores the interactions of these two important concepts: equilibria and limitations in learning. We introduce the question of whether equilibria continue to exist when agents have limitations. We look at the general effects limitations can have on agent behavior, and define a natural extension of equilibria that accounts for these limitations. Using this formalization, we make three major contributions: (i) a counterexample for the general existence of equilibria with limitations, (ii) sufficient conditions on limitations that preserve their existence, (iii) three general classes of games and limitations that satisfy these conditions. We then present empirical results from a specific multiagent learning algorithm applied to a specific instance of limited agents. These results demonstrate that learning with limitations is feasible, when the conditions outlined by our theoretical analysis hold.

## 1. Introduction

Multiagent domains are becoming more prevalent as more applications and situations require multiple agents. Learning in these systems is as useful and important as in single-agent domains, possibly more so. Optimal behavior in a multiagent system may depend on the behavior of the other agents. For example, in robot soccer, passing the ball may only be optimal if the defending goalie is going to move to block the player's shot and no defender will move to intercept the pass. This challenge is complicated by the fact that the behavior of the other agents is often not predictable by the agent designer, making learning and adaptation a necessary component of the agent itself. In addition, the behavior of the other agents, and therefore the optimal response, can be changing as they also adapt to achieve their own goals.

Game theory provides a framework for reasoning about these strategic interactions. The game theoretic concepts of stochastic games and Nash equilibria are the foundation for much of the recent research in multiagent learning, (e.g., Littman, 1994; Hu & Wellman, 1998; Greenwald & Hall, 2002; Bowling & Veloso, 2002). A Nash equilibrium defines a course of action for each agent, such that no agent could benefit by changing their behavior. So, all agents are playing optimally, given that the other agents continue to play according to the equilibrium.

From the agent design perspective, optimal agents in realistic environments are not practical. Agents are faced with all sorts of limitations. Some limitations may physically prevent certain behavior, e.g., a soccer robot that has traction limits on its acceleration. Other limitations are self-imposed to help guide an agent's learning, e.g., using a subproblem solution for advancing the ball down the field. In short, limitations prevent agents from playing optimally and possibly from following a Nash equilibrium.

This clash between the concept of equilibrium and the reality of limited agents is a topic of critical importance. Do equilibria exist when agents have limitations? Are there classes of domains or classes of limitations where equilibria are guaranteed to exist? This article introduces these questions and provides concrete answers. In particular, we introduce two models of limitations: *implicit games* and *restricted policy spaces*. We use these models to demonstrate in two key counterexamples the threat that limitations pose to the existence of equilibria. We refine a sufficient condition for the existence of equilibria, and use to prove existence in three classes of games and limitations. This analysis is peppered with examples of applications of these results. We conclude with an empirical example of learning with limitations, and a brief survey of related work.

The article is organized as follows. Section 2 introduces the stochastic game framework as a model for multiagent learning. We define the game theoretic concept of equilibrium, and examine the dependence of current multiagent learning algorithms on this concept. Section 3 enumerates and classifies some common agent limitations and presents two formal models incorporating the effects of limitations into the stochastic game framework. Section 4 is the major contribution of the article, presenting both proofs of existence for certain domains and limitations as well as counterexamples for others. Section 5 gives an example of how these results affect and relate to one particular multiagent learning algorithm. We present interesting results of applying a multiagent learning algorithm in a setting with limited agents. Section 6 compares our approach with other investigations of agent limitations in the fields of game theory and artificial intelligence. Finally, Section 7 concludes with implications of this work and future directions.

## 2. Stochastic Games

A *stochastic game* is a tuple $(n, \mathcal{S}, \mathcal{A}_{1...n}, T, R_{1...n})$, where,

- $n$ is the number of agents,

- $\mathcal{S}$ is a set of states,

- $\mathcal{A}_i$ is the set of actions available to agent $i$ with $\mathcal{A}$ being the joint action space, $\mathcal{A}_1 \times \ldots \times \mathcal{A}_n$,

- $T$ is a transition function, $\mathcal{S} \times \mathcal{A} \times \mathcal{S} \to [0, 1]$, such that,

$$\forall s \in \mathcal{S} \ \forall a \in \mathcal{A} \qquad \sum_{s' \in \mathcal{S}} T(s, a, s') = 1,$$

- and $R_i$ is a reward function for the $i$th agent, $\mathcal{S} \times \mathcal{A} \to \mathbb{R}$.

This framework is very similar to the framework of a Markov Decision Process (MDP). Instead of a single agent, though, there are multiple agents whose *joint action* determines the next state and rewards to the agents. The goal of an agent, as in MDPs, is to maximize its long-term reward. Notice, though, that each agent has its own independent reward function that it is seeking to maximize. The goal of maximizing "long-term reward" will be made formal in Section 2.2.

Stochastic games can also be thought of as an extension of the concept of matrix games to multiple states. Two common matrix games are in Figure 1. In these games there are two players; one selects a row and the other selects a column of the matrix. The entry of the matrix they jointly select determines the payoffs. Rock-Paper-Scissors in Figure 1(a) is a zero-sum game, where the column player receives the negative of the row player's payoff. In the general case (general-sum games; e.g., Bach or Stravinsky in Figure 1(b)) each player has an independent matrix that determines its payoff. Stochastic games, then, can be viewed as having a matrix game associated with each state. The immediate payoffs at a particular state are determined by the matrix entries $R_i(s, a)$. After selecting actions and receiving their rewards from the matrix game, the players are transitioned to another state and associated matrix game, which is determined by their joint action. So stochastic games contain both MDPs (when $n = 1$) and matrix games (when $|S| = 1$) as subsets of the framework.

$$R_r(s_0, \cdot) = \begin{pmatrix} 0 & -1 & 1 \\ 1 & 0 & -1 \\ -1 & 1 & 0 \end{pmatrix} \qquad R_r(s_0, \cdot) = \begin{pmatrix} 2 & 0 \\ 0 & 1 \end{pmatrix}$$

$$R_c(s_0, \cdot) = \begin{pmatrix} 0 & 1 & -1 \\ -1 & 0 & 1 \\ 1 & -1 & 0 \end{pmatrix} \qquad R_c(s_0, \cdot) = \begin{pmatrix} 1 & 0 \\ 0 & 2 \end{pmatrix}$$

(a) Rock-Paper-Scissors        (b) Bach or Stravinsky

Table 1: Two example matrix games.

## 2.1 Policies

Unlike in single-agent settings, deterministic policies, which associate a single action with every state, can often be exploited in multiagent settings. Consider Rock-Paper-Scissors as shown in Figure 1(a). If the column player were to play any action deterministically, the row player could win a payoff of one every time. This fact requires us to consider stochastic strategies and policies. A stochastic policy for player $i$, $\pi_i : \mathcal{S} \rightarrow PD(\mathcal{A}_i)$, is a function that maps states to mixed strategies, which are probability distributions over the player's actions. We use the notation $\Pi_i$ to be the set of all possible stochastic policies available to player $i$, and $\Pi = \Pi_1 \times \ldots \times \Pi_n$ to be the set of joint policies of all the players. We also use the notation $\pi_{-i}$ to refer to a particular joint policy of all the players except player $i$, and $\Pi_{-i}$ to refer to the set of such joint policies. Finally, the notation $\langle \pi_i, \pi_{-i} \rangle$ refers to the joint policy where player $i$ follows $\pi_i$ while the other players follow their policy from $\pi_{-i}$.

In this work, we make the distinction between the concept of stochastic policies and mixtures of policies. A mixture of policies, $\sigma_i : PD(\mathcal{S} \rightarrow \mathcal{A}_i)$, is a probability distribution over the set of deterministic policies. An agent following a mixture of policies selects a deterministic policy according to its mixture distribution at the start of the game and always follows this policy. This difference is similar to the distinction between mixed strategies and behavioral strategies in extensive-form games (Kuhn, 1953). Our work focuses on stochastic policies as they (i) are a more compact repre-sentation requiring $|A_i||S|$ parameters instead of $|A_i|^{|S|}$ parameters to represent the *complete* space of policies, (ii) are the common notion of stochastic policies in single-agent behavior learning, (e.g., Jaakkola, Singh, & Jordan, 1994; Sutton, McAllester, Singh, & Mansour, 2000; Ng, Parr, & Koller, 1999), and (iii) do not require the artificial commitment to a single deterministic policy at the start of the game, which can be difficult to understand within a learning context.

## 2.2 Reward Formulations

There are a number of possible reward formulations in single-agent learning that define the agent's notion of optimality. These formulations also apply to stochastic games. We will explore two of these reward formulations in this article: *discounted reward* and *average reward*. Although this work focuses on discounted reward, many of our theoretical results also apply to average reward.

**Discounted Reward.** In the discounted reward formulation, the value of future rewards is diminished by a discount factor $\gamma$. Formally, given a joint policy $\pi$ for all the agents, the value to agent $i$ of starting at state $s \in \mathcal{S}$ is,

$$V_i^\pi(s) = \sum_{t=0}^{\infty} \gamma^t E\left\{r_i^t | s^0 = s, \pi\right\},\tag{1}$$

where $r_i^t$ is the immediate reward to player $i$ at time $t$ with the expectation conditioned on $s$ as the initial state and the players following the joint policy $\pi$.

In our formulation, we will assume an initial state, $s_0 \in \mathcal{S}$, is given and define the goal of each agent $i$ as maximizing $V_i^\pi(s_0)$. This formulation differs from the usual goal in MDPs and stochastic games, which is to *simultaneously* maximize the value of all states. We require this weaker goal since our exploration into agent limitations makes simultaneous maximization unattainable.[1] This same distinction was required by Sutton and colleagues (2000) in their work on parameterized policies, one example of an agent limitation.

**Average Reward.** In the average reward formulation all rewards in the sequence are equally weighted. Formally, this corresponds to,

$$V_i^\pi(s) = \lim_{T \to \infty} \sum_{t=0}^{T} \frac{1}{T} E\left\{r_i^t | s^0 = s, \pi\right\},\tag{2}$$

with the expectation defined as in Equation 1. As is common with this formulation, we assume that the stochastic game is *ergodic*. A stochastic game is ergodic if for all joint policies any state can be reached in finite time from any other state with non-zero probability. This assumption makes the value of a policy independent of the initial state. Therefore,

$$\forall s, s' \in \mathcal{S} \qquad V_i^\pi(s) = V_i^\pi(s').$$

So any policy that maximizes the average value from one state maximizes the average value from all states. These results along with more details on the average reward formulation for MDPs are summarized by Mahadevan (1996).

---

1. This fact is demonstrated later by the example in Fact 5 in Section 4. In this game with the described limitation, if the column player randomizes among its actions, then the row player cannot simultaneously maximize the value of the left and right states.

For either formulation we will use the notation $V_i^\pi$ to refer to the value of the joint policy $\pi$ to agent $i$, which in either formulation is simply $V_i^\pi(s_0)$, where $s_0$ can be any arbitrary state for the average reward formulation.

## 2.3 Best-Response and Equilibria

Even with the concept of stochastic policies and well-defined reward formulations, there are, in general, still no optimal policies that are independent of the other players' policies. We can, though, define a notion of *best-response*.

**Definition 1** *For a game, the* best-response function *for player $i$,* $\mathrm{BR}_i(\pi_{-i})$*, is the set of all policies that are optimal given the other player(s) play the joint policy $\pi_{-i}$. A policy $\pi_i$ is optimal given $\pi_{-i}$ if and only if,*

$$\forall \pi_i' \in \Pi_i \qquad V_i^{\langle \pi_i, \pi_{-i} \rangle} \geq V_i^{\langle \pi_i', \pi_{-i} \rangle}.$$

The major advancement that has driven much of the development of game theory, matrix games, and stochastic games is the notion of a best-response equilibrium, or *Nash equilibrium* (Nash, Jr., 1950).

**Definition 2** *A* Nash equilibrium *is a joint policy,* $\pi_{i=1...n}$*, with*

$$\forall i = 1, \ldots, n \qquad \pi_i \in \mathrm{BR}_i(\pi_{-i}).$$

Basically, an equilibrium is a policy for each player where each is playing a best-response to the other players' policies. Hence, no player can do better by changing policies given that all the other players continue to follow the equilibrium policy. What makes the notion of an equilibrium interesting is that at least one, possibly many, exist in all matrix games and most stochastic games. This was proven by Nash (1950) for matrix games, Shapley (1953) for zero-sum discounted stochastic games, Fink (1964) for general-sum discounted stochastic games, and Mertens and Neyman (1981) for zero-sum average reward stochastic games. The existence of equilibria for general-sum average reward stochastic games is still an open problem (Filar & Vrieze, 1997).

In the Rock-Paper-Scissors example in Figure 1(a), the only equilibrium consists of each player playing the mixed strategy where all the actions have equal probability. In the Bach-or-Stravinsky example in Figure 1(b), there are three equilibria. Two consist of both players selecting their first action or both selecting their second. The third involves both players selecting their preferred cooperative action with probability 2/3, and the other action with probability 1/3.

## 2.4 Learning in Stochastic Games

Learning in stochastic games has received much attention in recent years as the natural extension of MDPs to multiple agents. The Minimax-Q algorithm (Littman, 1994) was the first reinforcement learning algorithm to explicitly consider the stochastic game framework. Developed for discounted reward, zero-sum stochastic games, the essence of the algorithm was to use Q-learning to learn the values of joint actions. The value of the next state was then computed by solving for the value of the unique Nash equilibrium of that state's Q-values. Littman and Szepesvari (1996) proved that under usual exploration requirements for both players, Minimax-Q would converge to the Nash equilibrium of the game, independent of the opponent's play beyond the exploration requirement.

Other algorithms have since been presented for learning in stochastic games. We will summarize these algorithms by broadly grouping them into two categories: *equilibrium learners* and *best-response learners*. The main focus of this summarization is to demonstrate how the existence of equilibria under limitations is a critical question to existing algorithms.

**Equilibrium Learners.** Minimax-Q has been extended in many different ways. Nash-Q (Hu & Wellman, 1998), Friend-or-Foe-Q (Littman, 2001), and Correlated-Q (Greenwald & Hall, 2002) are all variations on this same theme with different restrictions on the applicable class of games or the exact notion of equilibrium learned. All of the algorithms, though, seek to learn an equilibrium of the game directly, by iteratively computing intermediate equilibria. Some of the algorithms have theoretical guarantees of convergence to equilibrium play, others have empirical results to this effect. Like Minimax-Q, though, the policies learned are independent of the play of the other agents beyond exploration requirements. We refer collectively to these algorithms as *equilibrium learners*. What is important to observe is that these algorithms depend explicitly on the existence of equilibria. If an agent or agents were limited in such a way so that no equilibria existed then these algorithms would be, for the most part, ill-defined.[2]

**Best-Response Learners.** Another class of algorithms is the class of *best-response learners*. These algorithms do not explicitly seek to learn an equilibrium, instead seeking to learn best-responses to the other agents. The simplest example of one of these algorithms is Q-learning (Watkins, 1989). Although not an explicitly multiagent algorithm, it was one of the first algorithms applied to multiagent environments (Tan, 1993; Sen, Sekaran, & Hale, 1994). Another less naive best-response learning algorithm is WoLF-PHC (Bowling & Veloso, 2002), which varies the learning rate to account for the other agents learning simultaneously. Other best-response learners include Fictitious Play (Robinson, 1951; Vrieze, 1987), Opponent-Modeling Q-Learning (Uther & Veloso, 1997), Joint Action Learners (Claus & Boutilier, 1998), and any single-agent learning algorithm that learns optimal policies. Although these algorithms have no explicit dependence on equilibria, there is an important implicit dependence. If algorithms that learn best-responses converge when playing each other, then it must be to a Nash equilibrium (Bowling & Veloso, 2002). Therefore, all learning fixed points are Nash equilibria. In the context of agent limitations, this means that if limitations cause equilibria to not exist, then best-response learners could not converge.

This nonexistence of equilibria is exactly one of the problems faced by Q-learning in stochastic games. Q-learning is limited to deterministic policies. The deterministic policy limitation can, in fact, cause no equilibria to exist (see Fact 1 in Section 4.) So there are many games for which Q-learning cannot converge when playing with other best-response learners, such as other Q-learners.

In summary, both equilibrium and best-response learners depend, in some way, on the existence of equilibria. The next section explores agent limitations that are likely to be faced in realistic learning situations. In Section 4, we then present our main results examining the effect these limitations have on the existence of equilibria, and consequently on both equilibrium and best-response learners.

---

2. It should be noted that in the case of Minimax-Q, the algorithm and solution concept are still well-defined. A policy that maximizes its worst-case value may still exist even if limitations make it such that no equilibria exist. But, this minimax optimal policy might not necessarily be part of an equilibrium. Later, in Section 4, Fact 5, we present an example of a zero-sum stochastic game and agent limitations where the minimax optimal policies exist but do not comprise an equilibrium.

## 3. Limitations

The solution concept of Nash equilibrium depends on all the agents playing optimally. From the agent development perspective, agents have limitations that prevent this from being a reality. The working definition of limitation in this article is *anything that can restrict the agent from learning or playing optimal policies*. Broadly speaking, limitations can be classified into two categories: physical limitations and rational limitations. Physical limitations are those caused by the interaction of the agent with its environment and are often unavoidable. Rational limitations are limitations specifically chosen by the agent designer to make the learning problem tractable, either in memory or time. We briefly explore some of these limitations informally before presenting a formal model of limitations that attempts to capture their effect within the stochastic game framework.

### 3.1 Physical Limitations

One obvious physical limitation is that the agent simply is broken. A mobile agent may cease to move or less drastically may lose the use of one of its actuators preventing certain movements. Similarly, another agent may appear to be "broken" when in fact the motion is simply outside its capabilities. For example, in a mobile robot environment where the "rules" allow robots to move up to two meters per second, there may be a robot that is not capable of reaching that speed. An agent that is not broken, may suffer from poor control where its actions are not always carried out as desired, e.g., due to poorly tuned servos, inadequate wheel traction, or high system latency.

Another common physical limitation is hardwired behavior. Most agents in dynamic domains need some amount of hard-wiring for fast response and safety. For example, many mobile robot platforms are programmed to immediately stop if an obstacle is too close. These hardwired actions prevent certain behavior by the agent, which is often unsafe but is potentially optimal.

Sensing is a common area of agent limitations containing everything from noise to partial observability. Here we'll mention just one broad category of sensing problems: state aliasing. This occurs when an agent cannot distinguish between two different states of the world. An agent may need to remember past states and actions in order to properly distinguish the states, or may simply execute the same action in both states.

### 3.2 Rational Limitations

Rational limitations are a requirement for agents to learn in even moderately sized problems. Techniques for making learning scale, which often focus on near-optimal solutions, continue to be proposed and investigated in single-agent learning. They are likely to be even more necessary in multiagent environments which tend to have larger state spaces. We will examine a few specific methods.

In domains with sparse rewards one common technique is reward shaping, (e.g., Mataric, 1994). A designer artificially rewards the agent for actions the designer believes to be progressing toward the sparse rewards. These additional rewards can often speed learning by focusing exploration, but also can cause the agent to learn suboptimal policies. For example, in robotic soccer moving the ball down the field is a good heuristic for goal progression, but at times the optimal goal-scoring policy is to pass the ball backward to an open teammate.

Subproblem reuse also has a similar effect, where a subgoal is used in a portion of the state space to speed learning, (e.g., Hauskrecht, Meuleau, Kaelbling, Dean, & Boutilier, 1998; Bowling & Veloso, 1999). These subgoals, though, may not be optimal for the global problem and so prevent the agent from playing optimally. Temporally abstract options, either provided (Sutton, Precup, & Singh, 1998) or learned (McGovern & Barto, 2001; Uther, 2002), also enforce a particular sub-policy on a portion of the state space. Although in theory, the primitive actions are still available to the agents to play optimal policies, in practice abstraction away from primitive actions is often necessary in large or continuous state spaces.

Parameterized policies are receiving a great deal of attention as a way for reinforcement learning to scale to large problems, (e.g., Williams & Baird, 1993; Sutton et al., 2000; Baxter & Bartlett, 2000). The idea is to give the learner a policy that depends on far less parameters than the entire policy space actually would require. Learning is then performed in the smaller space of parameters using gradient techniques. Parameterized policies simplify and speed learning at the expense of possibly not being able to represent the optimal policy in the parameter space.

### 3.3 Models of Limitations

This enumeration of limitations shows that there are a number and variety of limitations with which agents may be faced, and they cannot be realistically avoided. In order to understand their impact on equilibria, we model limitations formally within the game theoretic framework. We introduce two models that capture broad classes of limitations: *implicit games* and *restricted policy spaces*.

**Implicit Games.** Limitations may cause an agent to play suboptimally, but it may be that the agent *is* actually playing optimally in a different game. If this new game can be defined within the stochastic game framework we call this the *implicit game*, in contrast to the original game called the *explicit game*. For example, reward shaping adds artificial rewards to help guide the agent's search. Although the agent is no longer learning an optimal policy in the explicit game, it is learning an optimal policy of some game, specifically the game with these additional rewards added to that agent's $R_i$ function. Another example is due to broken actuators preventing an agent from taking some action. The agent may be suboptimal in the explicit game, while still being optimal in the implicit game defined by removing these actions from the agent's action set, $A_i$. We can formalize this concept in the following definition.

**Definition 3** *Given a stochastic game* $(n, \mathcal{S}, \mathcal{A}_{1\ldots n}, T, R_{1\ldots n})$ *the tuple* $(n, \mathcal{S}, \hat{\mathcal{A}}_{1\ldots n}, \hat{T}, \hat{R}_{1\ldots n})$ *is an implicit game if and only if it is itself a stochastic game and there exist mappings,*

$$\tau_i : \mathcal{S} \times \hat{\mathcal{A}}_i \times \mathcal{A}_i \to [0, 1],$$

*such that,*

$$\forall s, s' \in \mathcal{S} \; \forall \hat{a}_i \in \hat{\mathcal{A}}_i \qquad \hat{T}(s, \langle \hat{a}_i \rangle_{i=1\ldots n}, s') = \sum_{a \in \mathcal{A}} \Pi_{i=1}^n \tau_i(s, \hat{a}_i, a_i) \; T(s, \langle a_i \rangle_{i=1\ldots n}, s').$$

*The $\tau_i$'s are mappings from the implicit action space into stochastic actions in the explicit action space.*

The $\tau_i$ mappings insure that, for each player, the choices available in the implicit game are a restriction of the choices in the explicit game. In other words, for any policy in the implicit game, there exists an equivalent, possibly stochastic policy, in the explicit game.

Reward shaping, broken actuators, and exploration can all be captured within this model. For reward shaping the implicit game is $(n, \mathcal{S}, \mathcal{A}_{1\ldots n}, T, \hat{R}_{1\ldots n})$, where $\hat{R}_i$ adds the shaped reward into the original reward, $R_i$. In this case the $\tau$ mappings are just the identity,

$$\tau_i(s, a, a') = \left\{ \begin{array}{ll} 1 & \text{if } a = a' \\ 0 & \text{otherwise} \end{array} \right. .$$

For the broken actuator example, let $a_i^0 \in \mathcal{A}_i$ be some null action for agent $i$ and let $a_i^b \in \mathcal{A}_i$ be some broken action for agent $i$ that under the limitation has the same effect as the null action. The implicit game, then, is $(n, \mathcal{S}, \mathcal{A}_{1...n}, \hat{T}, \hat{R}_{1...n})$, where,

$$\hat{T}(s, a, s') = \left\{ \begin{array}{ll} T(s, \langle a_i^0, a_{-i} \rangle, s') & \text{if } a_i = a_i^b \\ T(s, a, s') & \text{otherwise} \end{array} \right.$$

$$\hat{R}(s, a) = \left\{ \begin{array}{ll} R(s, \langle a_i^0, a_{-i} \rangle) & \text{if } a_i = a_i^b \\ R(s, a) & \text{otherwise} \end{array} \right. ,$$

and,

$$\tau_i(s, a) = \left\{ \begin{array}{ll} a_i^0 & \text{if } a = a_i^b \\ a & \text{otherwise} \end{array} \right. .$$

For $\epsilon$-exploration, as with broken actuators, only $\hat{T}$ and $\hat{R}$ need to be defined. They are just the $\epsilon$ combination of $T$ or $R$ with the transition probabilities or reward values of selecting a random action.

Limitations captured by this model can be easily analyzed with respect to their effect on the existence of equilibria. We now restrict ourselves to discounted reward stochastic games, where equilibria existence is known. Using the intuitive definition of an equilibrium as a joint policy such that "no player can do better by changing policies," an equilibrium in the implicit game achieves this definition for the explicit game. Since all discounted reward stochastic games have at least one equilibrium, so must the implicit game, which is also in this class. This equilibrium for the implicit game is then an equilibrium in the explicit game given that the agents are limited.

On the other hand, many of the limitations described above cannot be modeled in this way. None of the limitations of abstraction, subproblem reuse, parameterized policies, or state aliasing lend themselves to being described by this model. This leads us to our second, and in many ways more general, model of limitations.

**Restricted Policy Spaces.** The second model is that of *restricted policy spaces*, which models limitations as restricting the agent from playing certain policies. For example, an $\epsilon$-exploration strategy restricts the player to policies that select all actions with some minimum probability. Parameterized policy spaces have a restricted policy space corresponding to the space of policies that can be represented by their parameters. We can define this formally.

**Definition 4** *A restricted policy space for player $i$ is a non-empty and compact subset, $\overline{\Pi}_i \subseteq \Pi_i$, of the complete space of stochastic policies.*

The assumption of compactness[3] may at first appear strange, but it is not particularly limiting, and is critical for any equilibrium analysis.

---

3. Since $\overline{\Pi}_i$ is a subset of a bounded set, the requirement that $\overline{\Pi}_i$ is compact merely adds that the limit point of any sequence of elements from the set is also in the set.

| Physical Limitations | Implicit Games | Restricted Policies |
|---|:---:|:---:|
| Broken Actuators | ✓ | ✓ |
| Hardwired Behavior | ✓ | ✓ |
| Poor Control | ✓ | ✓ |
| State Aliasing | | ✓ |
| **Rational Limitations** | **Implicit Games** | **Restricted Policies** |
| Reward Shaping or Incentives | ✓ | |
| Exploration | ✓ | ✓ |
| State Abstraction/Options | | ✓ |
| Subproblems | | ✓ |
| Parameterized Policy | | ✓ |

Table 2: Common agent limitations. The column check-marks correspond to whether the limitation can be modeled straightforwardly using implicit games and/or restricted policy spaces.

It should be straightforward to see that parameterized policies, state aliasing (with no memory), and subproblem reuse, as well as exploration and broken actuators all can be captured as a restriction on policies the agent can play. Therefore they can be naturally described as restricted policy spaces. On the other hand, the analysis of the existence of equilibria under this model is not at all straightforward. Since restricted policy spaces capture most of the really interesting limitations we have discussed, this is precisely the focus of the rest of this article.

Before moving on to this analysis, we summarize our enumeration of limitations in Table 2. The limitations that we have been discussing are listed as well as denoting the model that most naturally captures their effect on agent behavior.

## 4. Existence of Equilibria

In this section we define formally the concept of a *restricted equilibrium*, which account for agents' restricted policy spaces. We then carefully analyze what can be proven about the existence of restricted equilibria. The results presented range from somewhat trivial examples (Facts 1, 2, 3, and 4) and applications of known results from game theory and basic analysis (Theorems 1 and 5) to results that we believe are completely new (Theorems 2, 3, and 4), as well as a critical counterexample to the wider existence of restricted equilibria (Fact 5). But all of the results are in a sense novel since this specific question has received no direct attention in the game theory nor the multiagent learning literature.

### 4.1 Restricted Equilibria

We begin by defining the concept of an equilibrium under the model of restricted policy spaces. First we need a notion of best-response that accounts for the players' limitations.

**Definition 5** *A restricted best-response for player $i$, $\overline{\mathrm{BR}}_i(\pi_{-i})$, is the set of all policies from $\overline{\Pi}_i$ that are optimal given the other player(s) play the joint policy $\pi_{-i}$.*

We can now use this to define an equilibrium.

**Definition 6** *A restricted equilibrium is a joint policy, $\pi_{i=1\ldots n}$, where,*

$$\pi_i \in \overline{\mathrm{BR}}_i(\pi_{-i}).$$

Hence, no player can do better by changing policies given that all the other players continue to follow the equilibrium policy, and the player can only switch to policies within their restricted policy space.

### 4.2 Existence of Restricted Equilibria

We can now state some results about when equilibria are preserved by restricted policy spaces, and when they are not. Unless otherwise stated (as in Theorems 2 and 4, which only apply to discounted reward), the results presented here apply equally to both the discounted reward and the average reward formulations. We will separate the proofs for the two reward formulations when needed. The first four facts show that the question of the existence of restricted equilibria does not have a trivial answer.

**Fact 1** *Restricted equilibria do not necessarily exist.*

**Proof.** Consider the Rock-Paper-Scissors matrix game with players restricted to the space of deterministic policies. There are nine joint deterministic policies, and none of these joint policies are an equilibrium. □

**Fact 2** *There exist restricted policy spaces such that restricted equilibria exist.*

**Proof.** One trivial restricted equilibrium is in the case where all agents have a singleton policy subspace. The singleton joint policy therefore must be a restricted equilibrium. □

**Fact 3** *If $\pi^*$ is a Nash equilibrium and $\pi^* \in \overline{\Pi}$, then $\pi^*$ is a restricted equilibrium.*

**Proof.** If $\pi^*$ is a Nash equilibrium, then we have

$$\forall i \in \{1 \ldots n\} \; \forall \pi_i \in \Pi_i \qquad V_i^{\pi^*} \geq V_i^{\left\langle \pi_i, \pi^*_{-i} \right\rangle}.$$

Since $\overline{\Pi}_i \subseteq \Pi_i$, then we also have

$$\forall i \in \{1 \ldots n\} \; \forall \pi_i \in \overline{\Pi}_i \qquad V_i^{\pi^*} \geq V_i^{\left\langle \pi_i, \pi^*_{-i} \right\rangle},$$

and thus $\pi^*$ is a restricted equilibrium. □

On the other hand, the converse is not true; not all restricted equilibria are of this trivial variety.

**Fact 4** *There exist non-trivial restricted equilibria that are neither Nash equilibria nor come from singleton policy spaces.*

**Proof.** Consider the Rock-Paper-Scissors matrix game from Figure 1. Suppose the column player is forced, due to some limitation, to play "Paper" exactly half the time, but is free to choose between "Rock" and "Scissors" otherwise. This is a restricted policy space that excludes the only Nash equilibrium of the game. We can solve this game using the implicit game model, by giving the limited player only two actions, $s_1 = (0.5, 0.5, 0)$ and $s_2 = (0, 0.5, 0.5)$, which the player can mix between. This is depicted graphically in Figure 1. We can solve the implicit game and convert the two actions back to actions of the explicit game to find a restricted equilibrium. Notice this restricted equilibrium is not a Nash equilibrium. □

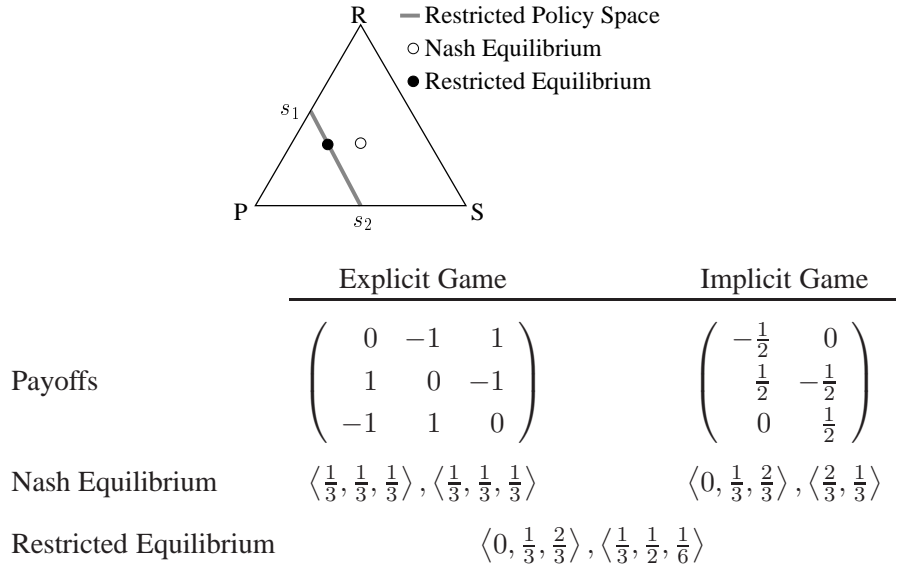

Figure 1: Example of a restricted equilibrium that is not a Nash equilibrium. Here, the column player in Rock-Paper-Scissors is restricted to playing only linear combinations of the strategies $s_1 = \left\langle \frac{1}{2}, \frac{1}{2}, 0 \right\rangle$ and $s_2 = \left\langle 0, \frac{1}{2}, \frac{1}{2} \right\rangle$.

Notice that the Fact 4 example has a convex policy space, i.e., all linear combinations of policies in the set are also in the set. Also, notice that the Fact 1 counterexample has a non-convex policy space. These examples suggest that restricted equilibria may exist as long as the restricted policy space is convex. The convexity restriction is, in fact, sufficient for matrix games.

**Theorem 1** *When $|S| = 1$, i.e. in matrix games, if $\overline{\Pi}_i$ is convex, then there exists a restricted equilibrium.*

The proof is based on an application of Rosen's theorem for concave games (Rosen, 1965) and is included in Appendix A. Surprisingly, the convexity restriction is not sufficient for stochastic games.

**Fact 5** *For a stochastic game, even if $\overline{\Pi}_i$ is convex, restricted equilibria do not necessarily exist.*

**Proof.** Consider the stochastic game in Figure 2. This is a zero-sum game where only the payoffs to the row player are shown. The discount factor $\gamma \in (0, 1)$. The actions available to the row player are $U$ and $D$, and for the column player $L$ or $R$. From the initial state, $s_0$, the column player may select either $L$ or $R$ which results in no rewards but with high probability, $1 - \epsilon$, transitions to the specified state (regardless of the row player's action), and with low probability, $\epsilon$, transitions to the opposite state. Notice that this stochasticity is not explicitly shown in Figure 2. In each of the resulting states the players play the matrix game shown and then deterministically transition back to the initial state. Notice that this game is unichain, where all the states are in a single ergodic set, thus satisfying the average reward formulation requirement. Also, notice that the game without limitations is guaranteed to have an equilibrium in both the discounted reward and the average reward, since it is zero-sum, formulations.

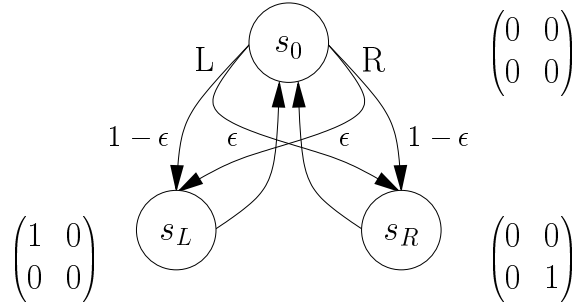

Figure 2: An example stochastic game where convex restricted policy spaces do not preserve the existence of equilibria.

Now consider the restricted policy space where players have to play their actions with the same probability in all states. So,

$$\overline{\Pi}_i \;=\; \left\{ \pi_i \in \Pi_i | \forall s, s' \in \mathcal{S} \; \forall a \in \mathcal{A} \quad \pi_i(s, a) = \pi_i(s', a) \right\}. \tag{3}$$

Notice that this is a convex set of policies. That is, if policies $x_1$ and $x_2$ are in $\overline{\Pi}_i$ (according to Equation 3), then for any $\alpha \in [0, 1]$, $x_3$ must also be in $\overline{\Pi}_i$, where,

$$x_3(s, a) = \alpha x_1(s, a) + (1 - \alpha) x_2(s, a). \tag{4}$$

The convexity of $\overline{\Pi}_i$ can be seen by examining $x_3(s', a)$ for any $s' \in \mathcal{S}$. From Equation 4, we have,

$$x_3(s', a) \;=\; \alpha x_1(s', a) + (1 - \alpha) x_2(s', a) \tag{5}$$
$$=\; \alpha x_1(s, a) + (1 - \alpha) x_2(s, a) \tag{6}$$
$$=\; x_3(s, a). \tag{7}$$

Therefore, $x_3$ is in $\overline{\Pi}_i$ and hence $\overline{\Pi}_i$ is convex.

This game, though, does not have a restricted equilibrium. The four possible joint deterministic policies, $(U, L)$, $(U, R)$, $(D, L)$, and $(D, R)$, are not equilibria. So if there exists an equilibrium it must be mixed. Consider any mixed strategy for the row player. If this strategy plays $U$ with probability less than $\frac{1}{2}$ then the only best-response for the column player is to play $L$; if greater than $\frac{1}{2}$ then the only best-response is to play $R$; if equal *then the only best-responses are to play L or R deterministically*. In all cases, all best-responses are deterministic, so this rules out mixed strategy equilibria, and so no equilibrium exists. □

Fact 5 demonstrates that convexity is not a strong enough property to guarantee the existence of restricted equilibria. The discussion in Section 2.4 concluded that the majority of existing multiagent learning algorithms hold some dependence on the existence of equilibria. Therefore, this result may have serious implications for the scaling of multiagent learning to large problems. We will explore these implications by first examining this example more closely, in order to identify a new sufficient condition. We will then seek to rescue the equilibrium notion by exploring specific situations where its existence can be preserved.

### 4.2.1 A SUFFICIENT CONDITION FOR EQUILIBRIA

Standard equilibrium proof techniques fail in the Fact 5 counterexample because the player's best-response sets are not convex, even though their restricted policy spaces are convex. Notice that the best-response to the row player mixing equally between actions is to play either of its actions deterministically. But, linear combinations of these actions (e.g., mixing equally) are not best-responses. This intuition is proven in the following lemma.

**Lemma 1** *For any stochastic game, if $\overline{\Pi}_i$ is convex and for all $\pi_{-i} \in \overline{\Pi}_{-i}$, $\overline{\mathrm{BR}}_i(\pi_{-i})$ is convex, then there exists a restricted equilibrium.*

The proof of this lemma relies on Kakutani's fixed point theorem using the convexity of the best-response as the critical condition for applying the theorem. Complete details of the proof can be found in Appendix B.

The consequence of this lemma is that, if we can prove that the sets of restricted best-responses are convex then restricted equilibria exist. Notice that the convexity of the restricted policy space is necessary for this condition (e.g., consider a uniform reward function resulting in $\overline{\mathrm{BR}}(\pi_{-i}) = \overline{\Pi}$). The Fact 5 counterexample demonstrates that it is not sufficient, since a restricted policy space can be convex without the restricted best-response sets being convex.

We now look at placing further restrictions either on the restricted policy spaces or the stochastic game, to insure provably convex best-response sets. After each existence proof for restricted equilibria, we give one or more practical examples of domains where the theoretical results apply. These results begin to enumerate a few general classes where restricted equilibria are guaranteed to exist.

### 4.2.2 A SUBCLASS OF RESTRICTED POLICIES

The first result for general stochastic games uses a stronger notion of convexity for restricted policy spaces.

**Definition 7** *A restricted policy space $\overline{\Pi}_i$ is* statewise convex*, if it is the Cartesian product over all states of convex strategy sets. Equivalently, if for all $x_1, x_2 \in \overline{\Pi}_i$ and all functions $\alpha : \mathcal{S} \to [0, 1]$, the policy $x_3(s, a) = \alpha(s)x_1(s, a) + (1 - \alpha(s))x_2(s, a)$ is also in $\overline{\Pi}_i$.*

In words, statewise convex policy spaces allow the policy's action probabilities at each state to be modified (within the restricted set) *independently*. We can show that this is a sufficient condition.

**Theorem 2** *In the discounted reward formulation, if $\overline{\Pi}_i$ is statewise convex, then there exists a restricted equilibrium.*

**Proof.** With statewise convex policy spaces, there exist optimal policies in the strong sense as mentioned in Section 2. Specifically, there exists a policy that can simultaneously maximize the value of all states. Formally, for any $\pi_{-i}$ there exists a $\pi_i \in \overline{\Pi}_i$ such that,

$$\forall s \in \mathcal{S} \ \forall \pi_i' \in \overline{\Pi}_i \qquad V_i^{\langle \pi_i, \pi_{-i} \rangle}(s) \geq V_i^{\langle \pi_i', \pi_{-i} \rangle}(s).$$

Suppose this were not true, i.e., there were two policies each of which maximized the value of different states. We can construct a new policy that in each state follows the policy whose value is larger for that state. This policy will maximize the value of both states that those policies maximized, and due to statewise convexity is also in $\overline{\Pi}_i$. We will use that fact to redefine optimality to this strong sense for this proof.

We will now make use of Lemma 1. First, notice the lemma's proof still holds even with this new definition of optimality. We just showed that under this redefinition, $\overline{\mathrm{BR}}_i(\pi_{-i})$ is non-empty, and the same argument for compactness of $\overline{\mathrm{BR}}_i(\pi_{-i})$ holds. So we can make use of Lemma 1 and what remains is to prove that $\overline{\mathrm{BR}}_i(\pi_{-i})$ is convex. Since $\pi_{-i}$ is a fixed policy for all the other players this defines an MDP for player $i$ (Filar & Vrieze, 1997, Corollary 4.2.11). So we need to show that the set of polices from the player's restricted set that are optimal for this MDP is a convex set. Concretely, if $x_1, x_2 \in \overline{\Pi}$ are optimal for this MDP, then the policy $x_3(s,a) = \alpha x_1(s,a) + (1-\alpha)x_2(s,a)$ is also optimal for any $\alpha \in [0,1]$. Since $x_1$ and $x_2$ are optimal in the strong sense, i.e., maximizing the value of all states simultaneously, then they must have the same per-state value.

Here, we will use the notation $V^x(s)$ to refer to the value of policy $x$ from state $s$ in this fixed MDP. The value function for any policy is the *unique* solution to the Bellman equations, specifically,

$$\forall s \in \mathcal{S} \qquad V^x(s) \;\; = \;\; \sum_a x(s,a) \left( R(s,a) + \gamma \sum_{s'} T(s,a,s')V^x(s') \right). \tag{8}$$

For $x_3$ then we get the following,

$$V^{x_3}(s) \;\; = \;\; \sum_a x_3(s,a) \left( R(s,a) + \gamma \sum_{s'} T(s,a,s')V^{x_3}(s') \right) \tag{9}$$

$$= \;\; \sum_a (\alpha x_1(s,a) + (1-\alpha)x_2(s,a)) \left( R(s,a) + \gamma \sum_{s'} T(s,a,s')V^{x_3}(s') \right) \tag{10}$$

$$= \;\; \alpha \sum_a x_1(s,a) \left( R(s,a) + \gamma \sum_{s'} T(s,a,s')V^{x_3}(s') \right) +$$

$$(1-\alpha) \sum_a x_2(s,a)) \left( R(s,a) + \gamma \sum_{s'} T(s,a,s')V^{x_3}(s') \right). \tag{11}$$

Notice that $V^{x_3}(s) = V^{x_1}(s) = V^{x_2}(s)$ is a solution to these equations, and therefore is the unique solution for the value of $x_3$. Therefore, $x_3$ has the same values as $x_1$ and $x_2$, and hence is also optimal. Therefore $\overline{\mathrm{BR}}_i(\pi_{-i})$ is convex, and from Lemma 1 we get the existence of restricted equilibria under this stricter notion of optimality, which also makes the policies a restricted equilibrium under our original notion of optimality, that is only maximizing the value of the initial state. $\square$

*Example: Multirobot Indoor Navigation.* Consider an indoor navigation domain with multiple robots traversing constrained hallways. The agents are the robots, all with their own navigational goals. A robot's state consists of a highly discretized representation of position and orientation resulting in an intractable number of states per robot. The global state is the joint states of all of the robots and is known by all the agents. The robots' actions are low-level navigation commands. We will construct an example of a useful restricted policy space for this domain that is statewise convex, and therefore satisfies the conditions of Theorem 2. Define a set of key decision points, $\mathcal{K}$, made up of the hallway intersections. These decision points define a subset of the complete state space, $\mathcal{K}^n \subseteq \mathcal{S}$. While between key decision points, each agent will use a fixed policy assigning probabilities to actions to reach the next decision point. The agents' policies are restricted so that decisions are only made when all agents have arrived at a key decision point, where they all simultaneously choose an action contingent on the global state. The agents can independently assign action probabilities for all states in $\mathcal{K}^n$, while all the action probabilities in $\mathcal{S} - \mathcal{K}^n$ are fixed. Hence, this is a statewise convex restricted policy space. Theorem 2 therefore applies and we conclude that there exists restricted equilibria in this domain.

Despite the above example, most rational limitations that allow reinforcement learning to scale are unlikely to be statewise convex restrictions. They often have a strong dependence between states. For example, parameterized policies involve far less parameters than the number of states, which is often intractably large. Either the majority of states have fixed action probabilities, as in the above example, or there will be fart too many parameters to optimize. Similarly, subproblems force whole portions of the state space to follow the same subproblem solution. Therefore, these portions of the state space do not select their actions independently. Abstraction and state aliasing use the same action probabilities across multiple states, and so also are not statewise convex. We now look at retaining a simple convexity condition on the restricted policy spaces, but examine constrained spaces of stochastic games.

### 4.2.3 SUBCLASSES OF STOCHASTIC GAMES

One way to relax from statewise convexity to general convexity is to consider only a subset of stochastic games. Theorem 1 is one example, where restricted equilibria exist for the subclass of matrix games with convex restricted policy spaces. Matrix games have a highly constrained "transition" function allowing only self transitions, which are independent of the players' actions. We can generalize this idea beyond single-state games.

**Theorem 3** *Consider no-control stochastic games, where all transitions are independent of the players' actions, i.e.,*

$$\forall s, s' \in \mathcal{S} \,\, \forall a, b \in \mathcal{A} \qquad T(s, a, s') = T(s, b, s').$$

*If $\overline{\Pi}_i$ is convex, then there exists a restricted equilibrium.*

**Proof (Discounted Reward).** This proof also makes use of Lemma 1, leaving us only to show that $\overline{\mathrm{BR}}_i(\pi_{-i})$ is convex. Just as in the proof of Theorem 2 we will consider the MDP defined for player $i$ when the other players follow the fixed policy $\pi_{-i}$. As before it suffices to show that for this MDP, if $x_1, x_2 \in \overline{\Pi}$ are optimal for this MDP, then the policy $x_3(s, a) = \alpha x_1(s, a) + (1 - \alpha) x_2(s, a)$ is also optimal for any $\alpha \in [0, 1]$.

Again we use the notation $V^x(s)$ to refer to the traditional value of a policy $x$ at state $s$ in this fixed MDP. Since $T(s, a, s')$ is independent of $a$, we can simplify the Bellman equations (Equation 8) to

$$V^x(s) = \sum_a x(s, a) R(s, a) + \gamma \sum_{s'} \sum_a x(s, a) T(s, a, s') V^x(s') \qquad (12)$$

$$= \sum_a x(s, a) R(s, a) + \gamma \sum_{s'} T(s, \cdot, s') V^x(s'). \qquad (13)$$

For the policy $x_3$, the value of state $s$ is then,

$$
\begin{aligned}
V^{x_3}(s) \;=\; & \alpha \sum_a x_1(s,a) R(s,a) + (1-\alpha) \sum_a x_2(s,a) R(s,a) + \\
& \gamma \sum_{s'} T(s,\cdot,s') V^{x_3}(s').
\end{aligned}
\tag{14}
$$

Using equation 13 for both $x_1$ and $x_2$ we get,

$$
\begin{aligned}
V^{x_3}(s) \;=\; & \alpha\left(V^{x_1}(s) - \gamma \sum_{s'} T(s,\cdot,s') V^{x_1}(s')\right) + \\
& (1-\alpha)\left(V^{x_2}(s) - \gamma \sum_{s'} T(s,\cdot,s') V^{x_2}(s')\right) + \\
& \gamma \sum_{s'} T(s,\cdot,s') V^{x_3}(s')
\end{aligned}
\tag{15}
$$

$$
\begin{aligned}
\;=\; & \alpha V^{x_1}(s) + (1-\alpha) V^{x_2}(s) + \\
& \gamma \sum_{s'} T(s,\cdot,s') \left(V^{x_3}(s') - \alpha V^{x_1}(s') - (1-\alpha) V^{x_2}(s')\right)
\end{aligned}
\tag{16}
$$

Notice that a solution to these equations is $V^{x_3}(s) = \alpha V^{x_1}(s) + (1-\alpha) V^{x_2}(s)$, and the Bellman equations must have a unique solution. Hence, $V^{x_3}(s_0)$ is equal to $V^{x_1}(s_0)$ and $V^{x_2}(s_0)$, which are equal since both are optimal. So $x_3$ is optimal, and $\overline{\mathrm{BR}}_i(\pi)$ is convex. Applying Lemma 1 we get that restricted equilibria exist. □

**Proof (Average Reward).** An equivalent definition to Equation 2 of a policy's average reward is,

$$
V_i^\pi = \sum_{s \in \mathcal{S}} d^\pi(s) \sum_a \pi(s,a) R(s,a),
\tag{17}
$$

where $d^\pi(s)$ defines the distribution over states visited while following $\pi$ after infinite time. For a stochastic game or MDP that is unichain we know that this distribution is independent of the initial state. In the case of no-control stochastic games or MDPs, this distribution becomes independent of the actions and policies of the players, and depends solely on the transition probabilities. So Equation 17 can be written,

$$
V_i^\pi = \sum_{s \in \mathcal{S}} d(s) \sum_a \pi(s,a) R(s,a).
\tag{18}
$$

As before, we must show that $\overline{\mathrm{BR}}_i(\pi_{-i})$ is convex to apply Lemma 1. Consider the MDP defined for player $i$ when the other players follow the policy $\pi_{-i}$. It suffices to show that for this MDP, if $x_1, x_2 \in \overline{\Pi}$ are optimal for this MDP, then the policy $x_3(s,a) = \alpha x_1(s,a) + (1-\alpha) x_2(s,a)$ is also optimal for any $\alpha \in [0,1]$. Using Equation 18, we can write the value of $x_3$ as,

$$
V_i^{x_3} \;=\; \sum_{s \in \mathcal{S}} d(s) \left( \sum_a x_3(s,a) R(s,a) \right)
\tag{19}
$$

$$
\;=\; \sum_{s \in \mathcal{S}} d(s) \left( \sum_a \left( \alpha x_1(s,a) + (1-\alpha) x_2(s,a) \right) R(s,a) \right)
\tag{20}
$$

$$= d(s) \left( \sum_a \alpha x_1(s,a)R(s,a) + \sum_a (1-\alpha)x_2(s,a)R(s,a) \right) \tag{21}$$

$$= \alpha \left( \sum_{s\in\mathcal{S}} d(s) \sum_a x_1(s,a)R(s,a) \right) + (1-\alpha) \left( \sum_{s\in\mathcal{S}} d(s) \sum_a x_2(s,a)R(s,a) \right) \tag{22}$$

$$= \alpha V_i^{x_1}(s) + (1-\alpha)V_i^{x_2}(s) \tag{23}$$

$$= \alpha V_i^{x_1}(s) + (1-\alpha)V_i^{x_1}(s) = V_i^{x_1}(s). \tag{24}$$

Therefore $x_3$ has the same average reward as $x_1$ and so is also optimal. So $\overline{\mathrm{BR}}_i(\pi_{-i})$ is convex and by Lemma 1 there exists an equilibrium. □

*Example: Space Probe.* Consider a space probe traveling on a predefined planetary trajectory. Consider using a multiagent approach to coordinating the gathering of data from scientific instruments. Each instrument is an agent making independent decisions and maximizing an assigned reward function. The state of the system is the position and orientation of the probe, which follows a fixed transition function. The state determines the effectiveness of the various data-gathering tasks (i.e., the agents' rewards), which also may be determined by the decisions of the other instruments. For example, camera images may be occluded when collecting atmospheric samples, or two instruments may draw too much power if activated simultaneously, causing both to fail. This is a no-control stochastic game, as the instruments' actions do not control the state, which is the predefined trajectory of the probe. By Theorem 3, restricted equilibria exist for all convex restricted policy spaces in this domain.

We can now merge Theorem 2 and Theorem 3 allowing us to prove existence of equilibria for a general class of games where only one of the player's actions affects the next state. Since this theorem's sufficient conditions are the most general of those presented, multiple examples of applications follow.

**Theorem 4** *Consider single-controller stochastic games (Filar & Vrieze, 1997), where all transitions depend solely on player 1's actions, i.e.,*

$$\forall s, s' \in \mathcal{S} \ \forall a, b \in \mathcal{A} \qquad a_1 = b_1 \Rightarrow T(s,a,s') = T(s,b,s').$$

*In the discounted reward formulation, if $\overline{\Pi}_1$ is statewise convex and $\overline{\Pi}_{i\neq 1}$ is convex, then there exists a restricted equilibrium.*

**Proof.** This proof again makes use of Lemma 1, leaving us to show that $\overline{\mathrm{BR}}_i(\pi_{-i})$ is convex. For $i = 1$ we use the argument from the proof of Theorem 2. For $i \neq 1$ we use the argument from Theorem 3. □

*Example: Haunted House Ride.* An example that helps to illustrate the concept of a single-controller stochastic game is an amusement park haunted house ride. Consider the system as consisting of multiple agents including the ride operator and the ride's passengers. Suppose the ride operator has choices over different ride routes therefore determining the state of the world for the passenger agents. The passengers, on the other hand, can only choose at each state where to look and when to try and scare their fellow passengers. Therefore, one agent (the operator) controls the transitions, while all the agents' (operator and passengers) affect the resulting rewards. Consider that the ride operator may be restricted to only making decisions at a few key decision points determined by the ride designer. Therefore, the operator's policy space is statewise convex. Hence, according to Theorem 4 if the passengers have convex restricted policy spaces resulting from their imposed limitations, then restricted equilibria exist.

*Example: Mobile Data Gathering.* The general pattern of the haunted house ride can also be seen in practical situations. Consider mobile data gathering, such as an extended version of the space probe example presented earlier. There is one navigational agent whose actions determine the state (e.g., position and orientation). The other agents are instruments choosing how to gather data at any given moment, as in the no-control version described above. The rewards are determined by all of the agents actions, but the state transition is defined only by the navigational agent. This fits the definition of a single-controller stochastic game and the Theorem holds for appropriate restricted policy spaces.

*Example: Asymmetric Competition* The above examples, although not strictly team games, do have a cooperative nature to them. Single-controller stochastic games can also be competitive. Consider an economic competition between two firms, where one firm's supply technology is primarily infrastructure-based, while the other's is very fluid. For example, a transportation company using rail lines versus a company using trucks. Since one agent's actions are implemented on a timescale far smaller than the other agent, we can effectively ignore that firm's state. The global state is just the built infrastructure of the first agent, and is only influenced by that agent's actions. The rewards, though, are determined by both actions of the agents, and may well be zero-sum. If the first firm uses a statewise convex restricted policy space (e.g., using key decision points) and the other firm uses a convex restricted policy space, then restricted equilibria exist.

The previous results have looked at stochastic games whose transition functions have particular properties. Our final theorem examines stochastic games where the rewards have a particular structure. Specifically we address team games, where the agents all receive equal payoffs.

**Theorem 5** *For team games, i.e.,*

$$\forall i, j \in \{1, \ldots, n\} \ \forall s \in \mathcal{S} \ \forall a \in \mathcal{A} \qquad R_i(s, a) = R_j(s, a),$$

*there exists a restricted equilibrium.*

**Proof.** The only constraints on the players' restricted policy spaces are those stated at the beginning of this section: non-empty and compact. Since $\overline{\overline{\Pi}}$ is compact, being a Cartesian product of compact sets, and player one's value in either formulation is a continuous function of the joint policy, then the value function attains its maximum (Gaughan, 1993, Corollary 3.11). Specifically, there exists $\pi^* \in \overline{\overline{\Pi}}$ such that,

$$\forall \pi \in \overline{\overline{\Pi}} \qquad V_1^{\pi^*} \geq V_1^{\pi}.$$

Since $\forall i V_i = V_1$ we then get that the policy $\pi^*$ maximizes all the players' rewards, and so each must be playing a restricted best-response to the others' policies. $\qquad \square$

*Example: Distributed Network Routing.* Consider a distributed network routing domain where message passing at nodes is controlled by an intelligent agent. The agent can only observe its own queue of incoming messages and congestion on outgoing links. Hence each agent has an aliased view of the global state. Therefore, policies mapping the observed state to actions are *convex* restricted policy spaces in the fully-observable stochastic game. All the agents seek to maximize global throughput. Hence, they have an identical reward function based on global packet arrival. As long as further agent limitations preserve the convexity of their policy spaces, by Theorem 5 there exists a restricted equilibrium.

### 4.3 Summary

Facts 1 and 5 provide counterexamples that show the threat limitations pose to equilibria. Theorems 1, 2, 3, 4, and 5 give us five general classes of stochastic games and restricted policy spaces where equilibria are guaranteed to exist. The fact that equilibria do not exist in general raises concerns about equilibria as a general basis for multiagent learning in domains where agents have limitations. On the other hand, combined with the model of implicit games, the presented theoretical results lays the initial groundwork for understanding when equilibria can be relied on and when their existence may be in question. These contributions also provide some formal foundation for applying multiagent learning in limited agent problems.

## 5. Learning with Limitations

In Section 2, we highlighted the importance of the existence of equilibria to multiagent learning algorithms. This section presents results of applying a particular learning algorithm to a setting of limited agents. We use the best-response learner, WoLF-PHC (Bowling & Veloso, 2002). This algorithm is rational, that is, it is guaranteed to converge to a best-response if the other players' policies converge and appropriate decayed exploration is used (Singh, Jaakkola, Littman, & Szepesvári, 2000). In addition, it has been empirically shown to converge in self-play, where both players use WoLF-PHC for learning. In this article we apply this algorithm in self-play to matrix games, both with and without player limitations. Since the algorithm is rational, if the players converge then their converged policies must be an equilibrium (Bowling & Veloso, 2002).

The specific limitations we examine fall into both the restricted policy space model as well as the implicit game model. One player is restricted to playing strategies that are the convex hull of a subset of the available strategies. From Theorem 1, there exists a restricted equilibrium with these limitations. For best-response learners, this amounts to a possible convergence point for the players. For the limited player, the WoLF-PHC algorithms were modified slightly so that the player maintains Q-values of its restricted set of available strategies and performs its usual hill-climbing in the mixed space of these strategies. The unlimited player is unchanged and *completely uninformed of the limitation of its opponent*. For all the experiments, we use very small learning rates and a large number of trials to better display how the algorithm learns and converges.

### 5.1 Rock-Paper-Scissors

The first game we examine is Rock-Paper-Scissors. Figure 3 shows the results of learning when neither player is limited. Each graph shows the mixed policy the player is playing over time. The labels to the right of the graph signify the probabilities of each action in the game's unique Nash equilibrium. Observe that the players' strategies converge to this learning fixed point.

Figure 4 shows the results of restricting player 1 to a convex restricted policy space, defined by requiring the player to play "Paper" exactly half the time. This is the same restriction as shown graphically in Figure 1. The graphs again show the players' strategies over time, and the labels to the right now label the game's restricted equilibrium, which accounts for the limitation (see Figure 1). The player's strategies now converge to this new learning fixed point. If we examine the expected rewards to the players, we see that the unrestricted player gets a higher expected reward in the restricted equilibrium than in the game's Nash equilibrium (1/6 compared to 0.) In summary, both players learn optimal best-response policies with the unrestricted learner appropriately taking advantage of the other player's limitation.

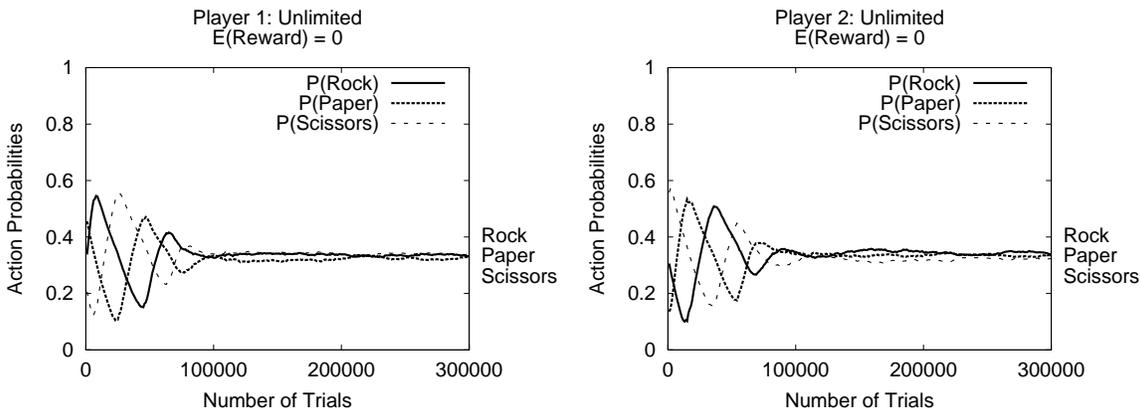

Figure 3: WoLF-PHC in Rock-Paper-Scissors. Neither player is limited.

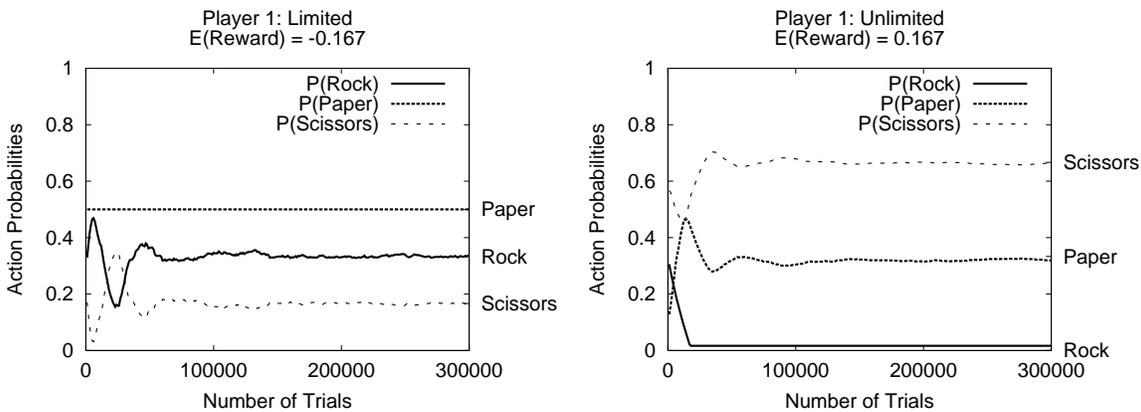

Figure 4: WoLF-PHC in Rock-Paper-Scissors. Player 1 must play "Paper" with probability $\frac{1}{2}$.

## 5.2 Colonel Blotto

The second game we examined is "Colonel Blotto" (Gintis, 2000), which is also a zero-sum matrix game. In this game, players simultaneously allot regiments to one of two battlefields. If one player allots more armies to a battlefield than the other, they receive a reward of one plus the number of armies defeated, and the other player loses this amount. If the players tie, then the reward is zero for both. In the unlimited game, the row player has four regiments to allot, and the column player has only three. The matrix of payoffs for this game is shown in Figure 5.

$$
R_r \; = \;
\begin{array}{c c}
 & \begin{array}{cccc} \textbf{3-0} & \textbf{2-1} & \textbf{1-2} & \textbf{0-3} \end{array} \\
\begin{array}{c} \textbf{4-0} \\ \textbf{3-1} \\ \textbf{2-2} \\ \textbf{1-3} \\ \textbf{0-4} \end{array} &
\left(\begin{array}{rrrr}
4 & 2 & 1 & 0 \\
1 & 3 & 0 & -1 \\
-2 & 2 & 2 & -2 \\
-1 & 0 & 3 & 1 \\
0 & 1 & 2 & 4
\end{array}\right)
\end{array}
\qquad
R_c \; = \;
\begin{array}{c c}
 & \begin{array}{cccc} \textbf{3-0} & \textbf{2-1} & \textbf{1-2} & \textbf{0-3} \end{array} \\
\begin{array}{c} \textbf{4-0} \\ \textbf{3-1} \\ \textbf{2-2} \\ \textbf{1-3} \\ \textbf{0-4} \end{array} &
\left(\begin{array}{rrrr}
-4 & -2 & -1 & 0 \\
-1 & -3 & 0 & 1 \\
2 & -2 & -2 & 2 \\
1 & 0 & -3 & -1 \\
0 & -1 & -2 & -4
\end{array}\right)
\end{array}
$$

Figure 5: Colonel Blotto Game.

Figure 6 shows experimental results with unlimited players. The labels on the right signify the probabilities associated with the Nash equilibrium to which the players' strategies converge. Player 1 is then given the limitation that it could only allot two of its armies, the other two would be allotted randomly. This is also a convex restricted policy space and therefore by Theorem 1 has a restricted equilibrium. Figure 7 shows the learning results. The labels to the right correspond to the action probabilities for the restricted equilibrium, which was computed by hand. As in Rock-Paper-Scissors, the players' strategies converge to the new learning fixed point. Similarly, the expected reward for the unrestricted player resulting from the restricted equilibrium is considerably higher than that of the Nash equilibrium ($0$ to $-14/9$), as the player takes advantage of the other's limitation.

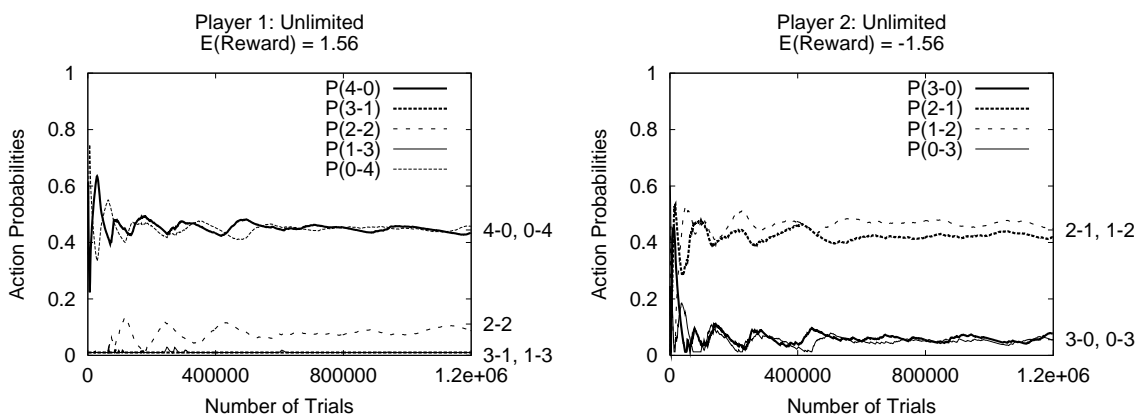

Figure 6: WoLF-PHC in Colonel Blotto. Neither player is limited.

There is one final observations about these results. In Section 3 we discussed the use of rational limitations to speed learning. Even in these very small single-state problems, our results demonstrate that limitations can be used to speed learning. Notice that convergence occurs more quickly in the limited situations where one of the players has less parameters and less freedom in its policy space. In the case of the Colonel Blotto game this is a dramatic difference. (Notice the x-axes differ by a factor of four!) In games with very large state spaces this will be even more dramatic. Agents will need to make use of rational limitations to do any learning at all, and similarly the less restricted agents will likely be able to benefit from taking advantage of the more limited learners.[4]

---

4. Strictly speaking, restricted agents are not always at a disadvantage. In zero-sum games, such as those for which results are presented here, and team-games where all players rewards are identical, restrictions can never benefit the agent. In general-sum games, such as Bach or Stravinsky from Table 1(b), this is not true. In this case, an imposed

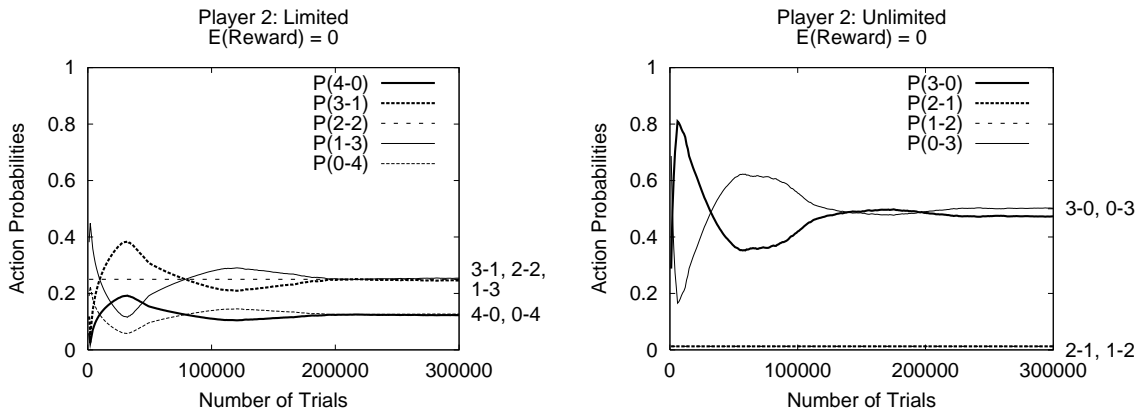

Figure 7: WoLF-PHC in Colonel Blotto. Player one is forced to randomly allot two regiments.

## 6. Related Work

It is commonly recognized that humans are not capable of being fully rational. There is a whole body of work in the social sciences, particularly economics and game theory, relating to understanding humans' *bounded rationality* (Simon, 1982). Bounded rationality is also a reality for artificial agents. As we discussed in Section 3, agent limitations are unavoidable in realistic problems. As a result, there has been much interesting work in both the fields of game theory and artificial intelligence on how limitations effect agent behavior.

A useful distinction presented by Simon (1976) is the difference between *substantive rationality* and *procedural rationality*. Substantive rationality is primarily concerned with whether a resulting behavior is appropriate for the achievement of specific goals given specific constraints. Procedural rationality is concerned with the behavior that is the result of appropriate deliberation. The difference is in whether it is the outcome that is appropriate to the goals or the deliberation that is appropriate to the goals.

The approach taken in this work is primarily procedural. We take the variety of well motivated learning procedures, both equilibrium learners and best-response learners, as given decision mechanisms. We also take the variety of mechanisms for making learning tractable (e.g., abstraction and approximation) as given mechanisms. The focus of our investigation is to see what is the resulting behavior of these combined mechanisms. In particular, do equilibria exist for equilibrium learners to learn and to which best-response learners can converge? This differs from other investigations which we'll briefly review here.

### 6.1 Game Theory

Much of the research in game theory on modeling bounded rationality, has focused on the investigation of procedural rationality (Rubinstein, 1998). This approach is crucial given the goal of modeling, explaining, and predicting *human behavior*. An example of this approach is that taken by Osborne and Rubinstein (1997). They define a decision procedure for mapping actions to discrete outcomes, rather than distributions over outcomes. This mapping is generated stochastically ac-

---

restriction, such as the row player being forced to select its first action, can actually improve the agent's equilibrium value (see also Littman & Stone, 2001; Gilboa & Samet, 1989).

cording to the actual distribution of outcomes for given actions. Given a mapping, the agent selects the action whose associated outcome is preferred. Given the decision procedure, they investigate whether an equilibrium fixed point exists and also how well such a procedure corresponds to human behavior in particular games.

Our work follows a similar approach of considering the outcome of using particular learning procedures. On the other hand, we are not dealing with the problem of human behavior but rather the behavior of artificial learning agents and how imposed limitations effect interactions between agents. By generalizing these effects into the models of implicit games and restricted policy spaces, we then ask similar questions. Do equilibrium fixed points exist and what is the expected outcome of employing such learning rules? This simplifies our problem since we're only analyzing the outcomes of known machine learning techniques rather than explaining human behavior.

In addition to the work along the lines of procedural rationality, there is a large body of traditional work following along the lines of substantive rationality. The particular goals of this approach are to model the agents' limitations in the game itself and find optimal behavior given the constraints of these limitations. These results include Bayesian games where other agents' utilities are uncertain a priori. Extensive form games with imperfect information is also an interesting model accounting for situations of state aliasing. Finally, strategies as finite state machines in repeated games try to account for limitations on computation and memory. A summary of this work can be found in standard game theory texts (e.g., Osborne & Rubinstein, 1994)).

## 6.2 Artificial Agents

While risking over generalizing, much of the work in the field of artificial agents that examines the effect of limitations on agent behavior has focused more on substantive rationality. In particular, a common thread of this work is the development of models of interaction and beliefs along with efficient algorithms for computing optimal or near-optimal behavior given these models. This approach focuses on what is appropriate behavior rather than what is the result of appropriate decision procedures. We will briefly consider some of the recent work that consider limitations and agent behavior, and give some comparison to the approach and results we have taken in this work.

### 6.2.1 COMPUTATIONAL LIMITATIONS

Larson and Sandholm (2001) consider the problem of reasoning in the presence of agents with limited computational resources. Specifically, they examine a two agent setting where agents have intractable individual problems, but may gain an improvement on their solutions by solving the also intractable joint problem. An agent's strategy determines how to invest its limited computational resources to improve its solution for its own problem, its opponent's problem, or the joint problem. When the deadline is reached, the agents must decide whether to simply use its own computed solution or bargain over a joint solution. Among other results, they demonstrate that the concept of Nash equilibrium can be extended to this setting, where the decision of computation is part of the agent's strategy. They call this concept *deliberation equilibria* and present algorithms for finding these equilibria solutions in certain situations.

The concept of deliberation equilibria has many similarities to the *restricted equilibria* concept we presented in Section 4. The core idea is to incorporate the agents' limitations into their choices of strategies. Restricted equilibria, though, incorporate limitations as restrictions on the agents' choices, rather than incorporating limitations by adding computational decisions into their choices.

This difference, though, is a fundamental one as the two approaches solve different problems. The complexity of decision-making in the Larson and Sandholm setting is directly due to the computational limitations placed on the agents. In Chapter 3, the complexity of decision-making is part of the task, and agent limitations are required in order to simplify the problem and make learning tractable. This core difference also explains the difference in results, since restricted equilibria are not guaranteed to exist in general.

### 6.2.2 GAME-TREE SEARCH WITH OPPONENT KNOWLEDGE

Jansen's Ph.D. thesis (1992) examined the issue of whether "optimal" play in the game-theoretic sense is in fact always "optimal". In the case of a fallible opponent that may make an incorrect evaluation of future states, the answer is naturally no. He proposed the idea of *speculative play*, which uses a prediction function for the opponent in determining its opponent's moves when searching a game-tree. The thesis focuses on deterministic, alternating-move, perfect information games, specifically a subgame of chess. He demonstrated that using speculative play with end-game databases as knowledge of the opponent has the potential for improving play against humans. In this work, limitations on human rationality are recognized, and so game-theoretic optimal play is discarded for optimization against particulars of human play.

### 6.2.3 PLANNING WITH OPPONENT MODELS.

Riley and Veloso (2000, 2002) present similar ideas for *using opponent models to make decisions* in a less abstract domain. They examine the problem of adapting to a specific opponent in simulated robotic soccer (Noda, Matsubara, Hiraki, & Frank, 1998). This work recognizes that modeling opponents is not a new idea and there has been a great deal of related research in the field of plan recognition. The work, though, contributes an actual technique for *using* opponent models to improve the team behavior, hence learning.

The team designer provides a set of opponent models of possible behavior of the other team. These models define a distribution of players' positions as a function of the ball's trajectory and their initial positions. During play, the actual opponent is matched to one of these models through Bayesian updating from actual observations. The model is then used to plan during set-play situations, such as throw-ins or goal-kicks. The planning uses hill-climbing on the ball-trajectory to find a trajectory with low probability of being intercepted by the other team *given the matched opponent model of their movement*. The planner then determines the positions of players and their actions so as to carry out the planned ball trajectory. This is compiled into a simple temporal network which can be executed in a distributed manner by the players. They demonstrate through controlled empirical experiments that the plans generated depend on the opponent model that is used during planning. The effectiveness of this technique in actual play against a real opponents, that may or may not match well with the models, is still unknown.

This technique allows a team of agents to adapt their behavior in a principled manner to better address a specific team. This does not strictly address the problem of concurrent learning in a multiagent setting, which is the primary focus of this work. The critical assumption is that the opponents' play is not changing over time. The authors mention this briefly and provide a small change to their Bayesian update through the addition of weight sharing. A small weight is added to each model to keep its probability bounded from zero. This effectively introduces a non-zero probability that the other team may change its behavior after each observation. There are no results, though, of the outcome of this approach if the other agents are also adapting. For example, it is not clear how robust the algorithm is to the other team employing a similar or identical approach to defending against a set-play. Will adaptation oscillate or is there an equilibrium point? How would such oscillation effect performance?

### 6.2.4 LIMITATIONS AS FINITE AUTOMATA.

Carmel and Markovitch (1996, 1997) examined a model-based approach to multiagent learning. They introduced algorithms for efficiently learning a model of the opponent as a deterministic finite automaton (DFA). Using this learned model, they then compute the optimal best response automaton. They examined this technique in the situation of repeated games and showed that their technique can learn effectively against a variety of DFA opponents. They also showed that their algorithm learned much more quickly (i.e., was more data efficient) than Q-learning, a model-free reinforcement learner. It is unknown how this compares with a model-based reinforcement learner such as prioritized sweeping (Moore & Atkeson, 1993).

This technique also does not address the problem of concurrent learning in a multiagent environment. Although a learning rule can be be modeled by a DFA, the number of states that a rule requires may be huge. For example, Q-learning with a decaying learning rate would be a completely intractable DFA. Also, the model-based learning agent is required to be strictly more powerful than its opponent, since its learning rule both learns the opponent's DFA and also generates its own. It is unclear whether any sort of equilibrium fixed point would be reached or even exists if two model-based DFA learners played each other. It is also not certain how DFA modeling scales to multiple state stochastic games.

### 6.2.5 SUMMARY

As we stated earlier, this body of work focuses primarily on substantive rationality: what is appropriate behavior given these assumptions? In addition, most of the approaches include the assumption of an asymmetric superiority over the other decision making agents. In the case of constructing DFA models of opponent behavior, there is an implicit assumption that the opponent behavior is simple enough to be constructed by the adapting agent. In the case of planning with opponent models, it is assumed that the opponent's behavior is not changing and so the learned model is accurate in predicting their future behavior. Although the work recognizes that other agents have limitations, they do so in a way that expects other agents to be strictly inferior. There are no results, empirical or theoretical, of applying these model-based techniques to situations of self-play, where all agents use the same adaptation procedure. Larson and Sandholm's work on computational limitations allows for limitations on the part of both agents. They do so by adding the complexity of computational decisions into the strategy space. Hence, their approach does not make this assumption.

## 7. Conclusion

Nash equilibrium is a crucial concept in multiagent learning both for algorithms that directly learn an equilibrium and algorithms that learn best-responses. Agent limitations, though, are unavoidable in realistic settings and can prevent agents from playing optimally or playing the equilibrium. In this article, we introduce and answer two critical questions: Do equilibria exist when agents have limitations? Not necessarily. Are there classes of domains or classes of limitations where equilibria are guaranteed to exist? Yes.

We have proven that for some classes of stochastic games and agent limitations equilibria are guaranteed to exist. We have also given counterexamples that help understand why equilibria do not exist in the general case. In addition to these theoretical results, we demonstrate the implications of these results in a real learning algorithm. We present empirical results that show that learning with limitations is possible, and equilibria under limitations is relevant.

There are two main future directions for this work. The first is continuing to explore the theoretical existence of equilibria. We have proven the existence of equilibria for some interesting classes of stochastic games and restricted policy spaces. We have also established in Lemma 1 a key criterion, the convexity of best-response sets, as the basis for further theoretical results. It is still unknown whether there are other general classes of games and limitations for which equilibria exist.

The second direction is the practical application of multiagent learning algorithms to real problems when agents have real limitations. The theoretical results we have presented and the empirical results on simple matrix games, give a foundation as well as encouraging evidence. There are still challenging questions to answer. How do specific limitations map on to the models that we explored in this article? Do equilibria exist for stochastic games and limitations typically faced in practice? What is the goal of learning when equilibria do not exist? This article lays the groundwork for exploring these and other important learning issues that are relevant to realistic multiagent scenarios.

## Acknowledgments

The authors are indebted to Martin Zinkevich for numerous insights as well as finding the example demonstrating Fact 5. The authors are grateful to Craig Boutilier and Tuomas Sandholm for helpful discussions. In addition, we owe gratitude to the team of anonymous reviewers for their careful eyes and invaluable comments.

This research was sponsored by the United States Air Force under Cooperative Agreements No. F30602-00-2-0549 and No. F30602-98-2-0135. The views and conclusions contained in this document are those of the authors and should not be interpreted as necessarily representing the official policies or endorsements, either expressed or implied, of the Defense Advanced Research Projects Agency (DARPA), the Air Force, or the US Government.

## Appendix A. Proof of Theorem 1

**Proof.** One might think of proving this theorem by appealing to implicit games as was used in Fact 4. In fact, if $\overline{\Pi}_i$ was a convex hull of a *finite* number of strategies, this would be the case. In order to prove it for any convex $\overline{\Pi}_i$ we apply Rosen's theorem about the existence of equilibria in concave games (Rosen, 1965). In order to use this theorem we need to show the following:

1. $\overline{\Pi}_i$ is non-empty, compact, and convex.

2. $V_i^\pi$ as a function over $\pi \in \overline{\Pi}$ is continuous.

3. For any $\pi \in \overline{\Pi}$, the function over $\pi_i' \in \overline{\Pi}_i$ defined as $V_i^{\left\langle \pi_i', \pi_{-i} \right\rangle}$ is concave.

Condition 1 is by assumption. In matrix games, where $\mathcal{S} = \{s_0\}$, we can simplify the definition of a policy's value from Equations 1 and 2. For discounted reward we get,

$$V_i^\pi = \frac{1}{1-\gamma} \sum_{a \in \mathcal{A}} R_i(s,a) \Pi_{i=1}^n \pi_i(s_0, a_i), \tag{25}$$

For average reward, the value is just the value of the one-shot matrix game given the players' policies. This is equivalent to setting $\gamma = 0$ in the Equation 25. Equation 25 shows that for both reward formulations the value is a multilinear function with respect to the joint policy and, therefore, is continuous. So Condition 2 is satisfied. Observe that by fixing the policies for all but one player, Equation 25 becomes a linear function over the remaining player's policy and so is also concave, satisfying Condition 3. Therefore Rosen's theorem applies and this game has a restricted equilibrium. □

## Appendix B. Proof of Lemma 1

**Proof.** The proof relies on Kakutani's fixed point theorem. We first need to show some facts about the restricted best-response function. First, remember that $\overline{\Pi}_i$ is non-empty and compact. Also, note that the value (with both discounted and average reward) to a player at any state of a joint policy is a continuous function of that joint policy (Filar & Vrieze, 1997, Theorem 4.3.7 and Lemma 5.1.4). Therefore, from basic analysis (Gaughan, 1993, Theorem 3.5 and Corollary 3.11), the set of maximizing (or optimal) points must be a non-empty and compact set. So $\overline{\mathrm{BR}}_i(\pi_{-i})$ is non-empty and compact.

Define the set-valued function, $F : \overline{\Pi} \to \overline{\Pi}$,

$$F(\pi) = \times_{i=1}^n \overline{\mathrm{BR}}_i(\pi_{-i}).$$

We want to show $F$ has a fixed point. To apply Kakutani's fixed point theorem we must show the following conditions to be true,

1. $\overline{\Pi}$ is a non-empty, compact, and convex subset of a Euclidean space,

2. $F(\pi)$ is non-empty,

3. $F(\pi)$ is compact and convex, and

4. $F$ is upper hemi-continuous.

Since the Cartesian product of non-empty, compact, and convex sets is non-empty, compact, and convex we have condition (1) by the assumptions on $\overline{\Pi}_i$. By the facts of $\overline{\mathrm{BR}}_i$ from above and the lemma's assumptions we similarly get conditions (2) and (3).

What remains is to show condition (4). Consider two sequences $x^j \to x \in \overline{\Pi}$ and $y^j \to y \in \overline{\Pi}$ such that $\forall j \ y^j \in F(x^j)$. It must be shown that $y \in F(x)$, or just $y_i \in \overline{\mathrm{BR}}_i(x)$. Let $v$ be $y_i$'s value against $x$. For the purposes of contradiction assume that there exists a $y_i'$ with higher value, $v'$, than $y_i$. Let $\delta = v' - v$. Since the value function is continuous, we can choose a point in the sequence where the policies are close enough to their limit points that they make an arbitrarily small change to the value. Specifically, let $N$ be large enough that the value of $y_i'$ against $x^N$ differs from $v'$ by at most $\delta/4$, and the value of $y_i$ against $x^N$ differs from $v$ by at most $\delta/4$, and the value of $y_i^N$ against $x^N$ differs from $y_i$ against $x^N$ by at most $\delta/4$. Adding all of these together, we have a point in the sequence $y_i^N$ whose value against $x^N$ is less than the value of $y_i'$ against $x^N$. So $y_i^N \notin \overline{\mathrm{BR}}_i(x^N)$, and therefore $y^N \notin F(x^N)$ creating our contradiction. The comparison of values of these various four joint policies is shown in Figure 8 and helps illustrate the resulting contradiction.

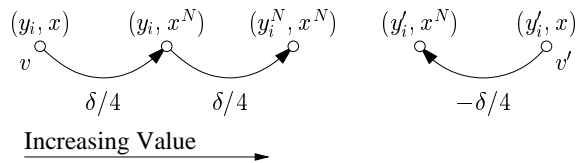

Figure 8: An illustration of the demonstration by contradiction that the best-response functions are upper hemi-continuous.

We can now apply Kakutani's fixed point theorem. So there exists $\pi \in \overline{\overline{\Pi}}$ such that $\pi \in F(\pi)$. This means $\pi_i \in \overline{\mathrm{BR}}_i(\pi_{-i})$, and therefore this is a restricted equilibrium. □